\documentclass[a4paper]{revtex4-1}
\usepackage{geometry}                

\usepackage{color}      
\usepackage{graphicx}
\usepackage{epstopdf}
\usepackage{bm}
\usepackage{amssymb, amsmath}
\usepackage{amsfonts}
\usepackage{amsthm}
\newcommand{\ket}[1]{|#1\rangle}

\begin{document}

\title{Observing wave/particle duality of light using topological charge }
\author{Olivier EMILE$^1$}\email[]{olivier.emile@univ-rennes1.fr}

 \author{Janine EMILE$^2$}

 \affiliation{$^1$Universit\'{e}  Rennes 1, 35042 Rennes cedex, France\\   $^2$UMR CNRS 6251 IPR, Universit\'{e}  Rennes 1, 35042 Rennes cedex, France\\  }
 
\date{\today}

\begin{abstract}
Wave particle duality, also called complementarity, is deeply rooted in the heart of quantum theory. It is fully exemplified in the famous Wheeler's delayed choice experiment where the choice of the wave nature (ability to interfere) or the particle like behavior (path distinguishability) is introduced a posteriori. We perform here a delayed choice experiment in a Mach-Zehnder interferometer, using a classical laser beam and twisted light in a given mode. We entangle the polarization and the twisted internal degrees of freedom, with the which-path-information external degree of freedom of the beam. The particle behavior of light arises from the quantization of the orbital angular momentum. It is demonstrated from torque and  light power measurements within $10 \%$ accuracy. We then experimentally evidence that the particle or wave behavior of light can be chosen a posteriori, even after the light has left the interferometer, at the moment of the detection.

\vspace{2cm}

\end{abstract}


\maketitle
\section*{Introduction}
\label{intro}
	The complementary discussion is at the very basis of quantum mechanics \cite{bohr1928,feynman1963,scully1991,bertet2001}. It is best evidenced in the so-called Wheeler's delayed choice experiment \cite{wheeler1978,wheeler1984,durr1998,legget2009,ma2016} that was first proposed for electrons. Briefly, as the output beam splitter of a Mach-Zehnder (MZ) interferometer is inserted, electrons in the two paths interfere. They behave as waves. When the beam splitter is removed, electrons travel either in one or the other arm. They behave as particles. Wheeler's claim is that, by delaying the choice for fixing the position of the final beam splitter, forces electrons to somehow decide whether to behave like particles or waves, long after they enter the interferometer \cite{hiley2006}. This has been experimentally demonstrated in optics using light from a single photon source \cite{Braig2003,jacques2007,tang2012} instead of electrons. Like for electrons, when the beam splitter is removed, photons travel either in one arm or in the other arm, and are detected at a single photon level. They behave as particles. When the beam splitter is inserted, photons interfere and behave as wave. Besides, since a long interferometric device is used, the choice can be made even after the light has already entered the interferometer. However, since a single photon source is used, the evidence of the wave nature of photons is not at all immediate. Indeed, the interference is due to a change in the optical path within the interferometer and its observation is constructed photon by photon and depends strongly on the statistics of the source. Moreover, the corpuscular nature of the light from the source has to be distinguished from the quantum nature of the detection.
	
	On the other hand, Electromagnetic (EM) fields carry angular momentum that may be detected via its transfer to matter \cite{beth1935,emile2014}. Within the paraxial approximation, two different forms of angular momentum may be distinguished. The first one is associated with the dynamical rotation of the direction of the fields around the propagation direction (Spin Angular Momentum, SAM), also called circular polarization. The second one is associated with the dynamical rotation of the light rays or the Poynting vector around the beam axis \cite{molina2007,barnett2017} (Orbital Angular Momentum, OAM). SAM is an intrinsic concept, that can be defined locally at a given position \cite{brown2011}. In particular, this means that two orthogonally polarized beams cannot interfere, even if only a small part of the beams is considered. On the contrary, OAM is defined from the whole beam, implying in particular that two OAM orthogonal beams can interfere \cite{courtial1998,vickers2008}. This has dramatic consequences in light entanglement experiments for example \cite{fickler2012}. It may also shine some new light in fundamental tests of quantum theory such as complementarity. Indeed, more specifically, could Wheeler's delayed choice experiment be carried out readily and unambiguously with a standard source, using OAM?

\renewcommand{\thesection}{\arabic{section}}
\renewcommand{\thesubsection}{\thesection .\arabic{subsection}}

\section{Experiment}

\begin{figure*}[htbp]
\centering 
\resizebox{1\textwidth}{!}{%
  \includegraphics{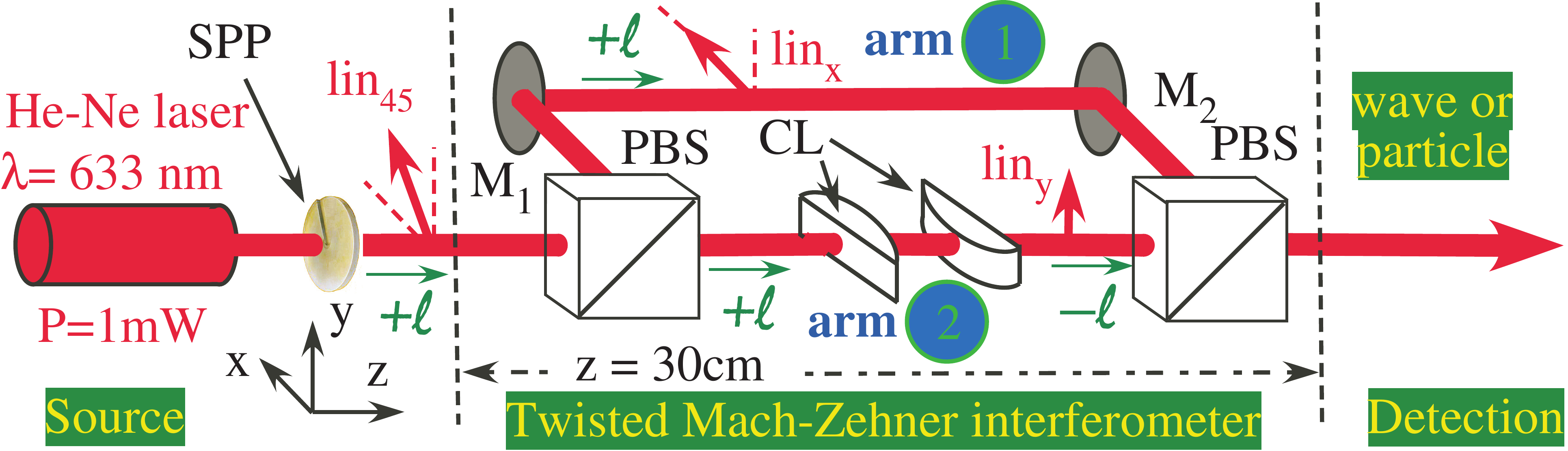}}

\caption{ Twisted MZ interferometer design. The fundamental mode of a helium-neon laser is transformed into a $\ell=+2$ LG mode with a SPP. Its polarization is at $45^\circ$ from the axes of the polarizing beam splitter (PBS). The PBS splits the beam in two parts. One part is transformed into a $\ell=-2$ LG mode thanks to two cylindrical lenses (CL), focal length $f=5$ cm. The two beams are then recombined with the help of a second PBS. M1, M2: mirrors.}
\label{fig1}
\end{figure*}

To address this issue, we have developed the experimental set-up described in figure \ref{fig1}. We use here the fundamental mode of a helium-neon laser (Melles Griot, $\lambda=633$ nm, P=1 mW) that is transformed in a higher order Laguerre Gaussian (LG) mode with a spiral phase plate (SPP) \cite{SPP}. The splitting of the beam is performed with a PBS set at $45^\circ$ from the polarization of the laser. The two arms of the MZ interferometer have the same intensity and correspond to linear vertical (along $y$) and horizontal (along $x$) polarizations of light respectively. The length of the interferometer is $z=30$ cm. 

The main difference from a usual MZ interferometer here, is that we use a LG laser beam at the entrance of the interferometer. Then, half of the light intensity travels in arm (1), with a polarization along $x$ in a LG mode with a topological charge or OAM $\ell=+2$. The other half travels in arm (2), with a polarization along $y$ in a LG mode transformed from $\ell=+2$ to $\ell=-2$ by two identical cylindrical lenses separated from twice their focal length \cite{padgett2002}. A second PBS then recombines the two arms that have orthogonal polarizations and conjugated topological charges. We carefully adjust mirrors M$_1$ and M$_2$ so that the two recombined beams overlap exactly. Note that the apparently longer length of arm (1), is compensated by the travel length inside the cylindrical glass lenses (CL), and also because the PBS are positioned  little off axis, so that the travel path of the beam running through arm (1) is longer in the PBS than for the beam running through arm (2). The interferometer is thus balanced.

We have entangled the Òwhich-pathÓ information (arm (1) or arm (2)) with the polarization (along $x$ or $y$) and the topological charge $+\ell$  or $-\ell$, of the beam. We made them non-separable. At the end of the interferometer the light beam is in a coherent superposition of (i) light that has travelled through arm (1), with a $\ell=+2$ topological charge and a polarization along $x$, and (ii) light that has travelled through arm (2), with a $\ell=-2$ topological charge and a polarization along $y$. In a more formal way, this coherent superposition of the EM field could be written (without normalization),
\begin{equation}
\label{eq1} 
\ket{\psi_{EM}^{MZ}}=\ket{1}\otimes\ket{+\ell}\otimes\ket{x}+\ket{2}\otimes\ket{-\ell}\otimes\ket{y}	, 	  
\end{equation} 	
the first item refers to the Òwhich-pathÓ information, either arm (1) or arm (2), the second item refers to the topological charge (OAM) of the beam and the last one to the polarization state (SAM). 
\begin{figure}[htbp]
\resizebox{1\textwidth}{!}{%
\includegraphics{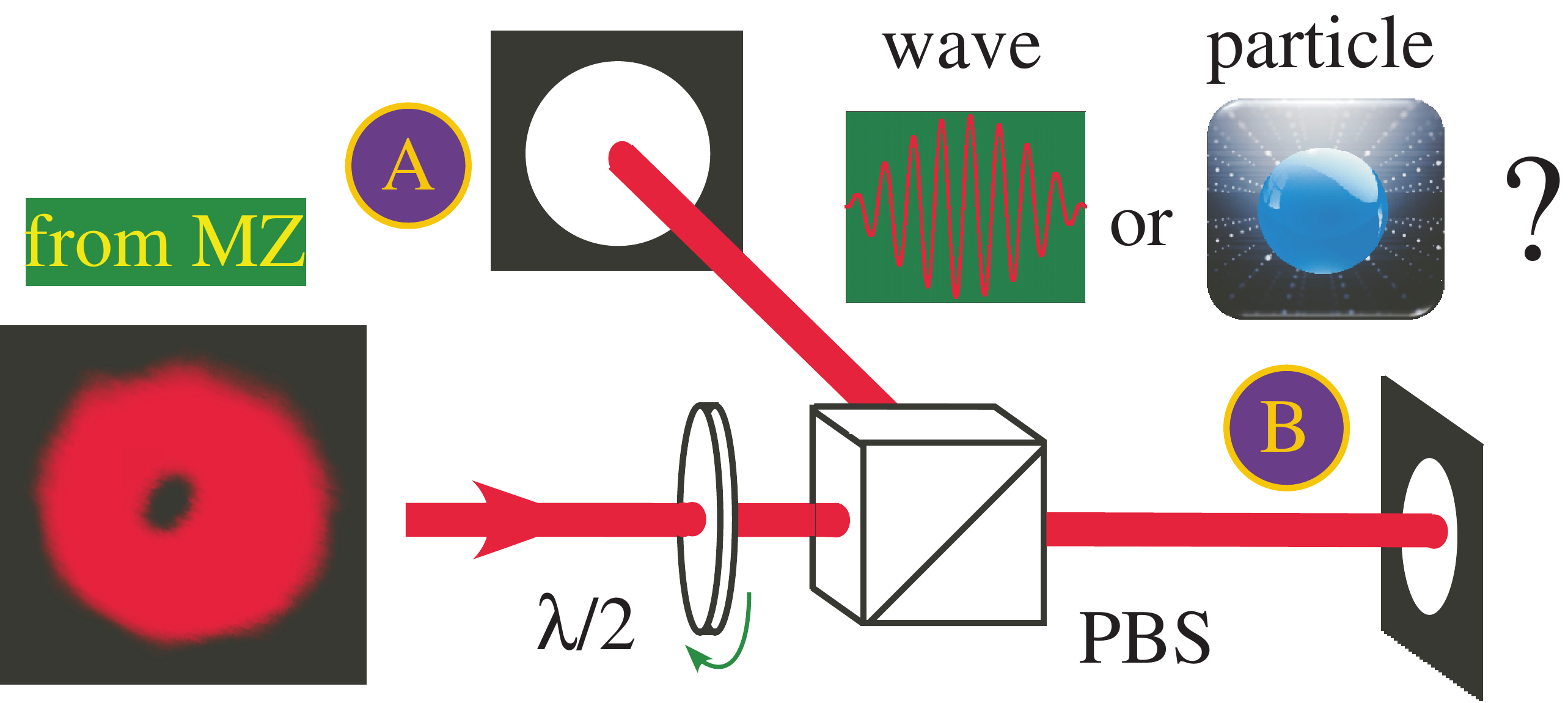}}
\caption{\label{fig2} Detection scheme. The light intensity from the MZ interferometer is split in two parts by a PBS, and observed with the naked eye, on two screens A and B. Depending on the orientation of the half wave plate ($\lambda/2$), light could behave as wave or particles. }
\end{figure}

\section{Results}
The detection scheme is displayed in figure \ref{fig2}. The light coming from the twisted MZ interferometer of figure 1 crosses a half wave plate ($\lambda/2$) that can be rotated and enters a PBS. Light is split in two parts and detected on two screens with the naked eye. 

\subsection{Particle behavior}
When the axes of the $\lambda/2$ are aligned with the axes of the PBS and correspond to the $x$ and $y$ directions, the light on A is polarized along $x$, has travelled through arm (1), and has a $\ell=+2$ topological charge. Conversely, the light on B is polarized along $y$, has travelled through arm (2), and has a $\ell=-2$ topological charge. In a more formal writing, the signal on screen A and B is the projection of equation (\ref{eq1}) on $\ket{x}$ and $\ket{y}$ respectively. It writes 
\begin{equation}
\label{eq2} 	
\ket{\psi_{EM}^{A}}=\ket{1}\otimes\ket{+\ell}\otimes\ket{x},	
\end{equation} 	
and
\begin{equation}
\label{eq3} 
\ket{\psi_{EM}^{B}}=\ket{2}\otimes\ket{-\ell}\otimes\ket{y}
\end{equation}
where $\ket{\psi_{EM}^{A}}$ and $\ket{\psi_{EM}^{B}}$ are the states of the light observed on the screens A and B respectively. Thus light that has travelled through arm (1) ends up in A, and light that has travelled through arm (2) ends up in B. It is as if, there was no beam splitter at the output of the interferometer in figure \ref{fig1} that recombines the beams. What do we detect experimentally? 

\begin{figure}[htbp]
\resizebox{1\textwidth}{!}{%
\includegraphics{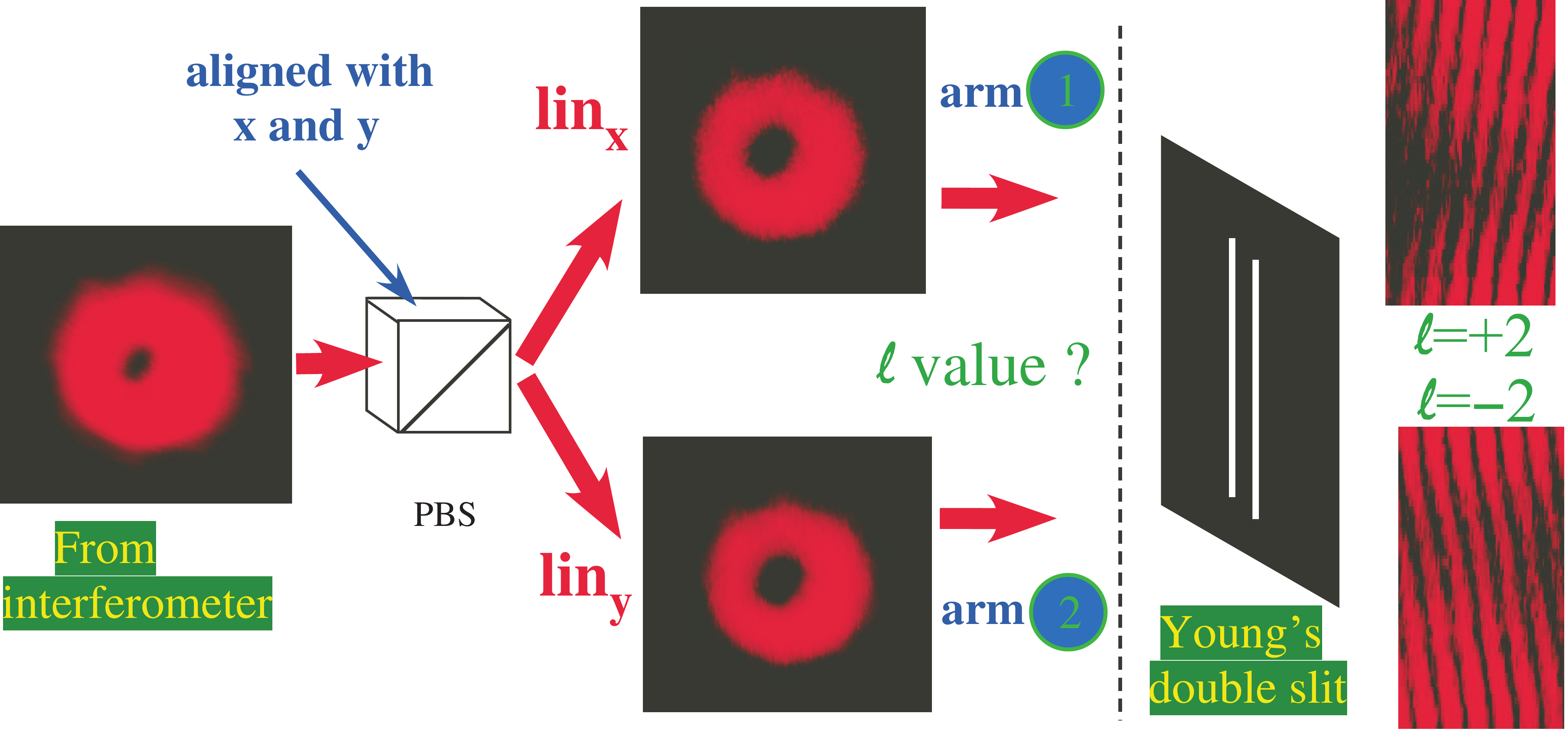}}
\caption{\label{fig3-0}  Measure of the topological charge of the beams after. 
The axis of the PBS correspond to the polarization of the light in the arms of the interferometer. Vortex beams are detected. Their topological charges are measured either with YoungÕs double slit \cite{emile2014b} or Fresnel Bi-prism \cite{emile2014c} technics. They are equal to $\ell=+2$ for arm (1) and $\ell=-2$ for arm (2) respectively.}
\end{figure}

\begin{figure}[htbp]
\resizebox{1\textwidth}{!}{%
\includegraphics{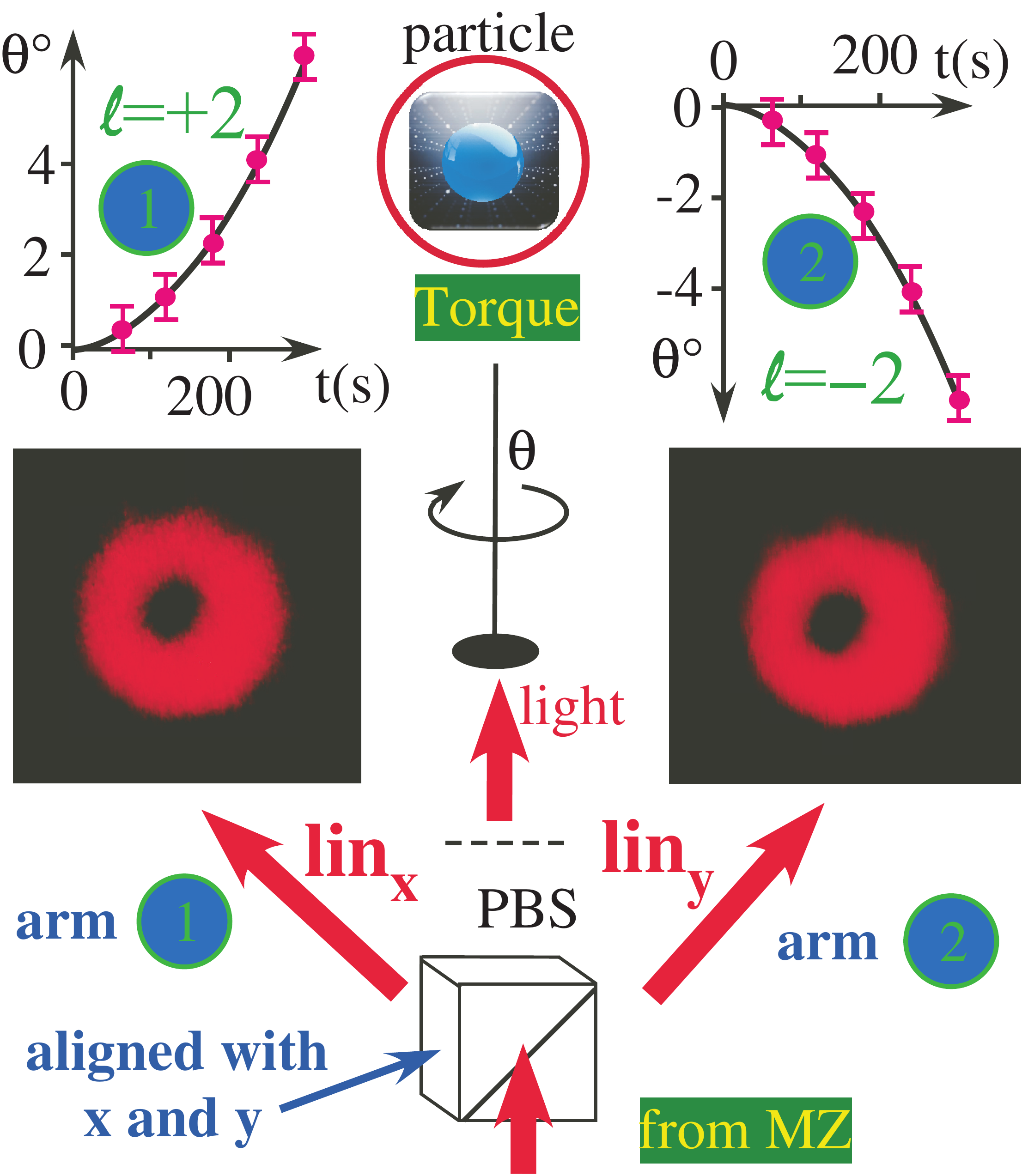}}
\caption{\label{fig3a}  Particle behavior. As the axes of the PBS correspond to the x and y directions, vortex beams are detected. Their topological charges equal to $\ell=+2$ and $\ell=-2$ and are measured by the torque deduced from the rotation of an absorbing disk.}
\end{figure}

The light intensities recorded on screens A and B look like the same. They have seemingly a vortex structure that may correspond to twisted OAM beams. We have probed their topological charge with Young's double slit experiment \cite{emile2014b}. The fringes after the slits correspond to $\ell=+2$ and $\ell=-2$ topological charges respectively (see  figure \ref{fig3-0}). The purity of the $\ell$ modes is better then $95\%$, from the interference pattern. This proves that we have experimentally entangled the which pass information with the polarization and the OAM state of light, as expected from equation (\ref{eq1}). On screen A, light has travelled through arm (1) is polarized along $x$ and has a topological charge $\ell=+2$, in agreement with equation (\ref{eq2}). Conversely, on screen B, light has travelled through arm (2) is polarized along $y$ and has a topological charge equal to $\ell=-2$, in agreement with equation (\ref{eq3}). Nevertheless, up to now, the results can be explained classically and do not prove the particle nature of light. Let us couple then the OAM with a macroscopic mechanical property. We have thus measured the available torque on both outputs of the PBS (see figure \ref{fig3a}).

The torque detection here is similar to what has been done for SAM \cite{emile2005}, or even for OAM in the case of radio emission \cite{emile2014,emile2016}. A 1.5 mm-diameter absorbing black paper (density 180 g.m$^{-2}$) hangs from a-10 cm-long ordinary cotton thread. The whole system is set in a vacuum chamber (pressure of 0.5 Pa). The beam coming out the PBS is focalized on the black paper with a 5-cm focal length ordinary lens. We register the rotation of the paper with a camera during 6 min (see figure \ref{fig3a}). We then measure the rotation angle that is in the ten degrees range. Since the rotation angle is small and the thread is long, the restoring torque is negligible. Besides, since the system is in a vacuum chamber, the friction is negligible either. The possible heating of the black paper has no influence on the torque. We observe a uniformly accelerated rotation. We then deduce the angular acceleration $\gamma$ that is equal to $\gamma=2.25\pm0.05\times10^{-6}$ rad.s$^{-2}$. We measure exactly the reversed acceleration for the beam at the other output of the PBS. 

We evaluate the moment of inertia $J$ of the piece of paper that is equal to $J=4.47\pm0.09\times10^{-14}$ kg.m$^2$. We then calculate the torque $\Gamma=J\gamma$. It equals  $\Gamma=1.00\pm0.05\times10^{-19}$ N.m for $\ell=+2$. The torque is reversed for $\ell=-2$. Assuming that all the light impinging on the paper is absorbed, due to angular momentum conservation, $\Gamma$ corresponds to the total OAM carried by the light. Before evacuating the chamber, we measure independently, at the same location the light power $P$ that is equal to $P=0.30\pm0.015$ mW. 

The ratio $\Gamma\omega/P$ equals $\Gamma\omega/P=2.01\pm 0.15$, $\omega$ being the light pulsation. Measuring only an integer number demonstrates that the OAM is quantized. We have checked that when we change the value of $\ell$ ($\ell=\pm1$,  $\pm2$, $\pm3$, $\pm4$), the ratio $\Gamma\omega/P$ still equals $\ell$. $\Gamma/2\hbar$ corresponds to the number of particles detected per second. Thus, because of the mechanical detection, independently of the photon statistics, neither on the nature of the light source, nor on the detection quantification, we detect particles carrying  $2\hbar$ OAM for $\ell=+2$ and  $-2\hbar$ OAM for $\ell=-2$, respectively, at an average rate of N$=\Gamma/2\hbar=29.5\times10^{14}$ particles per second. Unambiguously, light behaves as particles. Besides, the detection is performed with the naked eye, independently from any quantization nature consideration of the detector system. 

One may think that this result could be obtained classically. Indeed, if one considers a field carrying orbital angular momentum with a topological charge equal to $\ell$, following a straightforward "classical" calculation, one finds that $\Gamma\omega/P$ equals $\ell$. However, doing so, one implicitly considers that the topological charge is quantized. Because of cylindrical symmetry, it can only be equal an integer number. This reasoning here is very similar to what has been previously followed in the case of polarization \cite{Raman1931}, but also recently in the study of acoustic waves measuring both radiation pressure and torque \cite{Demore2012}.

Actually, from a mechanical point of view \cite{Truesdell1968}, angular momentum is a physical observable in its own right, in general independent of and not derivable from energy. The knowledge of one of these quantities doesn't imply the knowledge of the other. Energy on one side and orbital angular momentum on the other side are indeed truly independent quantities. 

\subsection{Wave behavior}
Let us rotate the $\lambda/2$ by $45^\circ$. The lights arriving on screens A and B are the projections of the light polarization on the eigen directions of the PBS, at 45$^\circ$ from $x$ and $y$. On screen A, it is the sum of part of the light that has travelled through arm (1), with a $\ell=+2$ topological charge and part of the light that has travelled through arm (2), with a $\ell=-2$ topological charge. In a more formal way, it is the projection of equation (\ref{eq1}) on $(\ket{x}+\ket{y})$
\begin{equation}
\label{eq4} 	
\ket{\psi_{EM}^{A}}=\ket{1}\otimes\ket{+\ell}+\ket{2}\otimes\ket{-\ell},	
\end{equation} 	
and the light on B is the difference of part of the light that has travelled through arm (1), with a $\ell=+2$ topological charge and part of the light that has travelled through arm (2), with a $\ell=-2$ topological charge. It is the projection of equation (\ref{eq1}) on $(\ket{x}-\ket{y})$. It then writes,
\begin{equation}
\label{eq5} 
\ket{\psi_{EM}^{B}}=\ket{1}\otimes\ket{+\ell}-\ket{2}\otimes\ket{-\ell}.
\end{equation}
However, as already stated, two conjugated LG modes interfere. Thus light that has travelled in arm (1) interferes with light that has travelled through arm (2). We shall detect interference patterns corresponding to conjugated LG beams\cite{emile2017}. The light then behaves as waves. 
	
\begin{figure}[htbp]
\resizebox{1\textwidth}{!}{%
\includegraphics{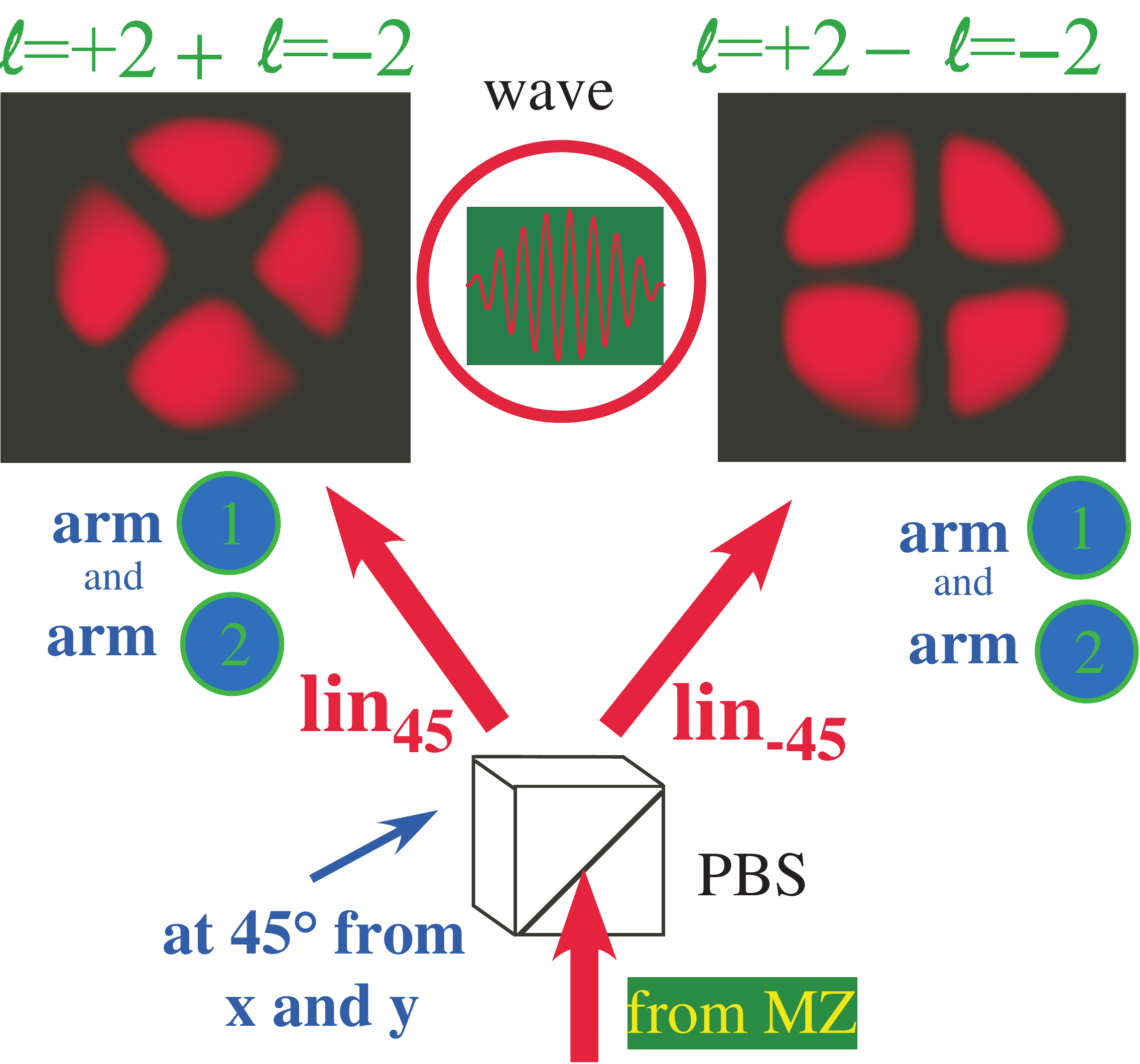}}
\caption{\label{fig3b}  Wave behavior. As the axes of the PBS are at $45^\circ$ from the $x$ and $y$ directions, $\ell=+2$  and $\ell=-2$ LG beams interfere on both output of the PBS, leading to 4 petals daisy flowers.}
\end{figure}

Figure \ref{fig3b} shows experimental pictures of the light intensity distribution detected on the screens A and B. They are 4 petals daisy flowers, as expected  from equations (\ref{eq4}) and (\ref{eq5}) and from the interference of conjugated LG beams. They are observed directly with the naked eye. The interference pattern on screens A and B are at $45^\circ$ from each other as expected. This also validates the use of equations (\ref{eq4}) and (\ref{eq5}) to describe the character of light. Of course the torque measurement in this configuration is equal to zero. The two macroscopic mechanical torques cancel, but light interferes. Finally, this clearly experimentally shows that, with this orientation of the $\lambda/2$, light behaves as wave. 

\section{Discussion}
	The choice to insert or remove the output beam splitter is made a posteriori. We have even increased the distance between the output of the interferometer and the detection scheme to $d=15$ m, in order to fulfill timing, also called sometimes  relativistic separation condition \cite{jacques2007}. We mean here that the choice of the switching must be space like separated from the entering of light in the interferometer. A fast electro-optical modulator (EOM) replaces the hand-rotated $\lambda/2$. It acts as an adjustable birefringent plate. When no voltage is applied, the axes of the PBS are aligned with $x$ and $y$. This situation corresponds to the removal of the beam splitter at the end of the interferometer in the original Wheeler's proposal. Experimentally, as in figure \ref{fig3a}, we measure the torque and the light power. We again find that light behaves as particles. 
	
	When a voltage, leading to a rotation of the polarization eigen-states by $45^\circ$, is applied, the axes of the PBS are at $45^\circ$ regarded to the polarization of the light in the arms of the interferometer. It is equivalent to the insertion of the beam splitter at the end of the interferometer in the original Wheeler's proposal. Experimentally, as in figure \ref{fig3b}, we observe four petals daisy flowers. We find that light, here also, behaves as waves. The switching time of the EOM can be as fast as 10 ns, corresponding to a 3-meter light propagation. This is much less than $d$. The switch can be random, thus fulfilling timing separation conditions. Of course, the fast switching is not evidenced with the naked eye. We use three fast photodiodes located on the same radius and rotated by $\pi/10$ from one another, regarded to the beam axis. If the three photodiodes measure the same intensity, it corresponds to figure \ref{fig3a}, whereas when an imbalanced between the photodiodes is measured, it corresponds to a daisy flower and to figure \ref{fig3b}.
	
	Besides, it has to be noted that, contrarily to what has been previously demonstrated \cite{jacques2007}, the choice to insert or remove the output beam splitter is performed not only after the light has entered the interferometer, but even long after the light has left the interferometer and the output beam splitter. It is performed just before the light detection. 
\begin{figure*}[htbp]
\resizebox{1\textwidth}{!}{%
\includegraphics{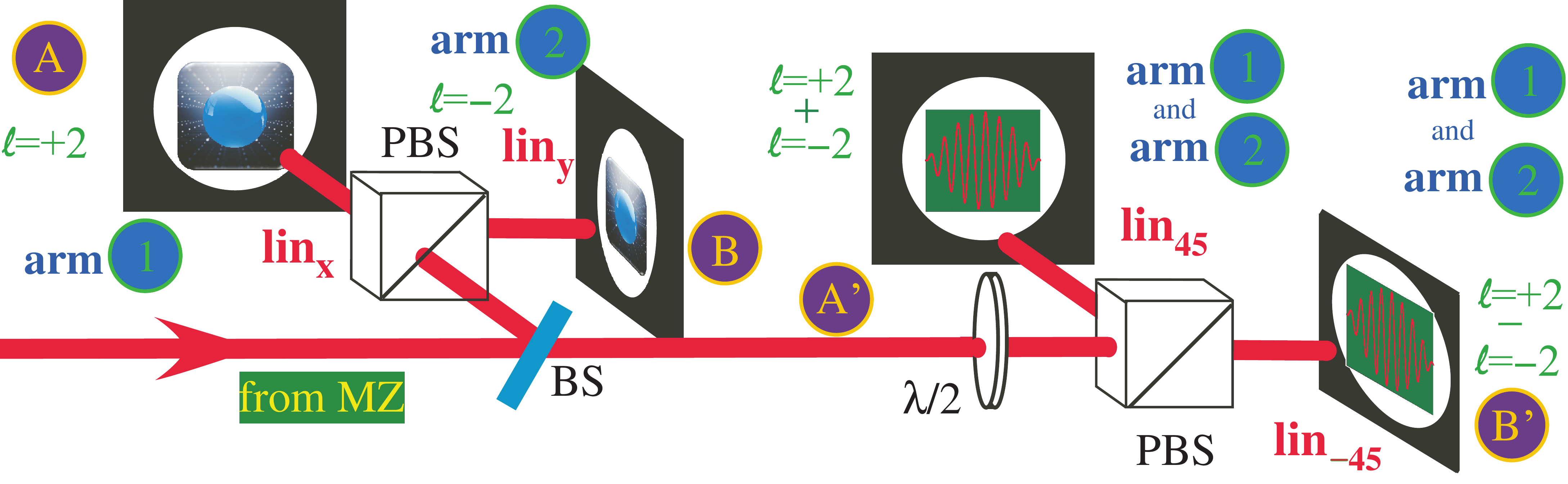}}
\caption{\label{fig4}  Complementarity test.  An ordinary beam splitter (BS) splits the light from the MZr interferometer in two equivalent parts. One part is again split in two parts A and B by a PBS which axis correspond to x and y directions, as in figure \ref{fig3a}, whereas the other part is split in two parts AÕ and BÕ by a PBS which polarization is at $45^\circ$ from the polarization of the arms of the MZ interferometer, as in figure \ref{fig3b}.}
\end{figure*}

Could we even demonstrate at the same time the particle and wave behavior using classical state of light? A similar question has been recently asked with a two-photon quantum source \cite{peruzzo2012,kaiser2012}. However, in their experiments, the authors observe an entity that is partially a wave and a particle. Here, we intend to probe the wave behavior and the particle behavior at the same time. We shall then use a single set-up to probe complementarity. Let us modify the detection scheme of figure \ref{fig2} now depicted in figure \ref{fig4}. Light first crosses an ordinary beam splitter. Part of the light is sent to a PBS which axes are aligned with the polarization of the light in the arms of the interferometer. The detection A or B probes which way of the interferometer the light passes through. It is related to the particle like nature of light. It leads to the same results as in  figure \ref{fig3a}.

 The other part of the light is sent to an other PBS which axes are at $45^\circ$ compared with the polarization of the light in the arms of the interferometer. It probes the wave nature of light. It leads to the same result as in figure \ref{fig3b}. Thus, at the same time, within the same interferometer, part of the light behaves as particles, whereas, simultaneously, the other part of the neighboring light behaves as wave that interferes.

\section*{Conclusion}
We have performed a Wheeler's type experiment using a classical He-Ne laser and conjugated OAM beams in the arms of the interferometer. The wave behavior arises from the interference of the two paths of the modified interferometer, as in usual MZ interferometer with light. The particle behavior arises from the OAM that is indeed quantized. Moreover, we have answered the question of when the decision whether light will display wave like or particle like behavior \cite{hiley2006} is made. As recently demonstrated for atoms \cite{manning2015}, the decision here for light is made at the point of measurement. Moreover, this decision is independent from the behavior of the neighboring light at the same time, at the same location, in the same interferometer. 

These counterintuitive results may then lead to deny the physical reality of the wavefunction to maintain causality. However, a recent work has directly verified the physical reality of the wavefunction \cite{Zhou2017}. It turns out that our results are in agreement with this recent work since, at the output of the interferometer, the nature of light is a vector vortex state, i.e. the state described by its wavefunction, that collapses at the moment of the measurement.

\textbf{Acknowledgments.} The authors thank Jean Ren\'e THEBAULT for technical assistance.

\end{document}